\documentclass[11pt]{article}

\def\mytitle#1{\setcounter{equation}{0}
\setcounter{footnote}{0}
\begin{flushleft}\Large\textbf{#1}\end{flushleft}
\vspace{0.25cm}}
\def\myname#1{\leftline{{\large #1}}\vspace{-0.13cm}}
\def\myplace#1#2{\small\begin{flushleft}\textit{#1}\\
\texttt{#2}\end{flushleft}}
\newenvironment{contribution}{\normalsize\noindent}{}
\def\myclassification#1{\small\noindent
Pacs no :
       #1\vspace{0.5cm}}
\usepackage{graphicx}% Include figure files
\begin{document}

\mytitle{Gravitational Collapse in Generalized Vaidya Space-Time
for Lovelock Gravity Theory}

\vskip0.2cm \myname{ Prabir Rudra~$*$} \vskip0.2cm \myname{
Ritabrata Biswas~$\dag$} \vskip0.2cm \myname{ Ujjal Debnath~$*$}

\myplace{$*$Department of Mathematics, Bengal Engineering and
Science University, Shibpur, Howrah-711 103, India.}
{[prabirrudra92@gmail.com, ujjal@iucaa.ernet.in,
ujjaldebnath@yahoo.com]}

\myplace{$\dag$ Department of Mathematics, Jadavpur University,
Kolkata-700 032, India.} {[biswas.ritabrata@gmail.com]}

\myclassification{04.20.Dw, 04.20.Ex, 04.20.Cv, 04.70.Bw}

\begin{abstract}
In this work, we have assumed the generalized Vaidya solution in
Lovelock theory of gravity in $(n+2)$-dimensions. It has been
shown that Gauss-Bonnet gravity, dimensionally continued Lovelock
gravity and pure Lovelock gravity can be constructed by suitable
choice of parameters. We have investigated the occurrence of
singularities formed by the gravitational collapse in above three
particular forms of Lovelock theory of gravity. The dependence of
the nature of singularity on the existence of radial null geodesic
for Vaidya space-time has been specially considered. In all the
three models, we have shown that the nature of singularities
(naked singularity or black hole) completely depend on the
parameters. Choices of various parameters are shown in tabular
form. In Gauss-Bonnet gravity theory, it can be concluded that the
possibility of naked singularity increases with increase in
dimensions. In dimensionally continued Lovelock gravity, the naked
singularity is possible for odd dimensions for several values of
parameters. In pure Lovelock gravity, only black hole forms due to
the gravitational collapse for any values of parameters. It has
been shown that when accretion is taking place on a collapsing
object, it is highly unlikely to get a black hole. Finally on
considering the phantom era in the expanding universe it is
observed that there is no possibility of formation of a black hole
if we are in the Gauss-Bonnet gravity considering the accreting
procedure upon a collapsing object.
\end{abstract}

\section{Introduction}

Gravitational collapse is one of the most important problem in
classical general relativity. The study of gravitational collapse
was started by Oppenheimer and Snyder (1939). They studied
collapse of dust with a static Schwarzschild exterior while
interior space-time is represented by Friedman like solution. One
would like to know whether, and under what initial conditions,
gravitational collapse results in black hole (BH) formation. In
particular, one would like to know if there are physical collapse
solutions that lead to naked singularities (NS). In last few
years, there have been extensive studies on gravitational collapse
in order to investigate the nature of the singularities. The study
of gravitational collapse of spherically symmetric space-times led
to many examples (Eardly et al, 1979; Christodoulou, 1984; Neuman,
1986; Waugh et al, 1986, Dwivedi et al, 1989) of NSs. There is no
general theory of the nature and visibility of singularities.
There do exist a number of exact solutions of the Einstein
equation which admit, depending upon the initial data, BHs or NSs
(Joshi, 1993, 2000; Clarke, 1993; Wald, 1997; Jhingan, 1999;
Singh, 1999). In particular the Vaidya solution (Vaidya, 1951)
(contains outgoing radiation) is extensively used to show that the
end state of collapse for regular initial data results in a NS.\\

Harko et al (2000) have studied the gravitational collapse of
strange matter and analyzed the condition for formation of a NS in
the spherically symmetric Vaidya space-time. It has been shown
that depending on the initial distribution of density and velocity
and on the constitutive nature of the collapsing matter, either a
BH or a NS is formed. Santos and collaborators (Santos, 1984,
1985; de Oliveira et al, 1985, 1986, 1987, 1988) included the
dissipation in the source by allowing radial heat flow (while the
body undergoes radiating collapse). Ghosh and Deskar (2000, 2003)
have considered collapse of a radiating star with a plane
symmetric boundary and have concluded with some general remarks.
Ghosh et al (2002) have discussed the study of collapse of
radiating star in Vaidya space-time in $(n+2)$-dimensions. Wang et
al (1999) has generalized the Vaidya solution which include most
of the known solutions to the Einstein equation such as
anti-de-Sitter charged Vaidya solution. Husain solution has been
used to study the formation of a BH with short hair (Brown et al,
1997) and can be considered as a generalization of Vaidya solution
(Wang et al, 1999). Recently, Patil et al (2005, 2006) have
studied the gravitational collapse of the Husain solution in four
and five dimensional space-times and Debnath et al (2008) have
studied the gravitational collapse of the Husain solution in
$(n+2)$-dimensional space-times for various
types of equation of state.\\

The Lovelock gravity theory (Lovelock, 1971) is a generalization of Einstein
gravity theory in higher dimensional space-times, but the Lovelock
theory is not a higher derivative gravity theory. The Lagrangian
of the Lovelock gravity consists of the dimensionally extended
Euler densities

\begin{equation}\label{1}
{\cal L}=\sum_{i=0}^{p}c_{i}{\cal L}_{i}
\end{equation}

where $p\le [(n+1)/2]$ (where, $[x]$ denotes the integer part of
the number $x$), $c_{i}$ are arbitrary constants with dimension of
$[length]^{2i-2}$, $(n+2)$ is the space-time dimension and ${\cal
L}_{i}$ are the Euler densities of a $(2i)$-dimensional manifold
(Cai et al, 2008)

\begin{equation}\label{2}
{\cal
L}_{i}=\frac{1}{2^{i}}~\sqrt{-g}~\delta^{a_{1}...a_{i}b_{1}...b_{i}}_{c_{1}...c_{i}d_{1}...d_{i}}~
R^{c_{1}d_{1}}_{a_{1}b_{1}}~...~R^{c_{i}d_{i}}_{a_{i}b_{i}}
\end{equation}

Here, the generalized delta function is totally antisymmetric in
both sets of indices. ${\cal L}_{0}$ is set to one, therefore the
constant $c_{0}$ is just the cosmological constant. ${\cal L}_{1}$
gives us the usual scalar curvature term. If we set $c_{1}=1$,
${\cal L}_{2}$ gives just the Gauss-Bonnet (GB)term. So the
Einstein-Gauss-Bonnet(EGB) gravity is a special case of Lovelock's
theory of gravitation, whose Lagrangian just contains the first
three terms in (1).\\

Static spherically symmetric BH solutions can be found in Lovelock
theory in the sense that a metric function is determined by
solving for a real root of a polynomial equation (Myers et al,
1988; Wheeler, 1986). More recently, static, non-spherically
symmetric BH solutions have been also found in the Lovelock
gravity (Cai, 2004). Spherically symmetric BH solutions in the GB
gravity have been found and discussed in (Myers et al, 1988;
Whiler, 1986; Boulware et al, 1985 ) and rotating GB BHs have been
discussed in (Kim et al, 2008). Some exact solutions for a
Vaidya-like solution in the EGB gravity have been found in
(Kobayashi, 2005; Maeda, 2006; Dominguez et al, 2006; Ghosh et al,
2008). Gravitational collapse and study of NS formation in
dimensionally continued Lovelock gravity have been
investigated in ref.(Nozawa et al, 2006; Ilha et al, 1997, 1999).\\

In this paper, we are mainly studying the nature of singularities
(BH or NS) formed by the gravitational
collapse in Lovelock theory of gravity. In section 2, we present
the brief overview of generalized Vaidya solution in general
Lovelock theory of gravity. Next we investigate the gravitational
collapse in GB gravity, dimensionally continued Lovelock
gravity and pure Lovelock gravity in sections 3-5. We have
discussed a accretion phenomena upon the collapsing object
in these gravity theories in section 6. Finally, the
paper ends with a short discussions in section 7.\\

\section{Brief Overview of Generalized Vaidya Solution in Lovelock \\ Gravity Theory}

The metric ansatz in $(n+2)$-dimensional spherically symmetric
Vaidya space-time can be written as

\begin{equation}\label{3}
ds^{2}=-f(v,~r)dv^{2}+2dvdr+r^{2}d\Omega_{n}^{2}
\end{equation}

where $r$ is the radial coordinate ($0<r<\infty$) and the null
coordinate $v$ ($-\infty<v<\infty$) stands for advanced Eddington
time coordinate and $f(v,~r)=1-\frac{m(v,~r)}{r^{n-1}}$ where
$m(v,~r)$ gives the gravitational mass inside the sphere of radius
$r$ and $d\Omega_{n}^{2}$ is the line element of a $n$-dimensional
unit sphere. The energy-momentum tensor for the Vaidya null
radiation in the space-time is given by $T_{ab}=\mu l_{a}l_{b}$,
where $l_{a}=(1,0,0,...,0)$ and $\mu$ is the energy density of the
null radiation. Recently, Cai et al (2008) have calculated the
field equations for Vaidya metric and their solutions in Lovelock
gravity. Here we write briefly the field equations and solutions
from their work.\\

The Einstein's tensors for Vaidya metric in Lovelock gravity are
given by [Cai et al, 2008]

\begin{equation}\label{4}
{\cal G}_{v}^{v}={\cal
G}_{r}^{r}=\frac{1}{2}\sum_{i=0}^{p}c_{i}~\frac{n!}{(n-2i+1)!}\left(\frac{1-f}{r^{2}}
\right)^{i} \left[(n-2i+1)-\frac{i~rf'}{(1-f)} \right]
\end{equation}

\begin{equation}\label{5}
{\cal
G}_{v}^{r}=-\frac{1}{2}\sum_{i=0}^{p}c_{i}~\frac{n!}{(n-2i+1)!}\left[\left(\frac{i\dot{f}}{r}
\right) \left(\frac{1-f}{r^{2}} \right)^{i-1}\right]
\end{equation}

\begin{eqnarray*}\label{6}
{\cal
G}_{k}^{j}=\frac{1}{2}\delta_{k}^{j}\sum_{i=0}^{p}c_{i}~\frac{(n-1)!}{(n-2i+1)!}\left(\frac{1-f}{r^{2}}
\right)^{i-1}
 \left[(n-2i)(n-2i+1)\frac{1-f}{r^{2}}-2(n-2i+1)\frac{f'}{r}-if''\right.
\end{eqnarray*}
\begin{equation}\label{7}
\left.~~~~~~~~~~~~~~~~~~~~~ ~~~~~~~~~~~~~~~~~~~~~
~~~~~~~~~~~~~~~~~~~~~~~~~~~~~~~~~~~~~~~~~~~~
+i(i-1)\left(\frac{rf'}{1-f}\right)^{2} \right]
\end{equation}

The Einstein's field equation is given by

\begin{equation}\label{8}
{\cal G}_{ab}=8\pi G T_{ab}
\end{equation}

Since we have ${\cal G}_{r}^{r}={\cal G}_{v}^{v}$, so from above
equation (7) we must have $T_{r}^{r}=T_{v}^{v}$. Now we assume that the
spherical part of the energy-momentum tensor has the form
$T_{i}^{i}=\sigma T_{r}^{r}=\sigma T_{v}^{v}$,
where $\sigma$ is a constant.\\

Now from conservation equation $\nabla_{a}T^{a}_{b}=0$, we get the
following two equations as [Cai et al, 2008]

\begin{equation}\label{9}
\partial_{v}T_{v}^{v}+\partial_{r}T^{r}_{v}+\frac{n}{r}T^{r}_{v}=0
\end{equation}
and
\begin{equation}\label{10}
\partial_{r}T_{r}^{r}+\frac{n(1-\sigma)}{r}T^{r}_{r}=0
\end{equation}

For pure null radiation, we get $T_{r}^{r}=T_{v}^{v}=0$. Here we
consider $T_{r}^{r}=T_{v}^{v}\ne 0$. So from equation (9), we get
[Cai et al, 2008]

\begin{equation}\label{11}
T_{r}^{r}=T_{v}^{v}=C(v)r^{-n(1-\sigma)}
\end{equation}

where $C(v)$ is a function of $v$. Now define a function

\begin{equation}\label{12}
F(v,~r)=\frac{1-f(v,~r)}{r^{2}}
\end{equation}

From (4), (7), (10) and (11), we get the equation

\begin{equation}\label{13}
\frac{1}{2}\sum_{i=0}^{p}c_{i}~\frac{n!}{(n-2i+1)!}~\partial_{r}(r^{n+1}F^{i})=8\pi
G C(v)~r^{-n\sigma}
\end{equation}

which integrates to yield

\begin{equation}\label{14}
\sum_{i=0}^{p}c_{i}~\frac{n!}{(n-2i+1)!}~F^{i}=16\pi G
\left(\frac{m(v)}{\Omega_{n}r^{n+1}} +
\frac{C(v)\Theta(r)}{r^{n+1}}\right)
\end{equation}

where $m(v)$ is arbitrary function of $v$,
$\Omega_{n}=\frac{\pi^{\frac{n}{2}}}{\Gamma(1+\frac{n}{2})}$ and
$\Theta(r)=\int r^{n\sigma}dr$ i.e.,

\begin{equation}\label{15}
\Theta(r)=\left\{\begin{array}{c}
\nonumber\frac{r^{n\sigma+1}}{n\sigma+1}~~~~when~\sigma\ne -\frac{1}{n}\\\\
\ln r~~~~when~\sigma=-\frac{1}{n}
\end{array}\right.
\end{equation}

Now from equations (5), (7) and (11) , we get

\begin{equation}
\frac{1}{2}\sum_{i=0}^{p}c_{i}~\frac{n!~r}{(n-2i+1)!}~\partial_{v}F^{i}=8\pi
G T^{r}_{v}
\end{equation}

Now using (13) and (15), we obtain

\begin{equation}
T_{v}^{r}=\mu= \frac{\dot{m}(v)}{\Omega_{n}r^{n+1}} +
\frac{\dot{C}(v)\Theta(r)}{r^{n+1}}
\end{equation}

We see that (10) and (16) satisfy (8). From (6), (7) and (11), we
get (using $T_{k}^{j}=0$)

\begin{equation}
\delta_{k}^{j}\sum_{i=0}^{p}c_{i}~\frac{(n-1)!}{(n-2i+1)!}~\frac{1}{r^{n-1}}
~\partial_{rr}(r^{n+1}F^{i})=0
\end{equation}

So the energy-momentum tensor can be written as

\begin{equation}
T_{ab}=\mu l_{a}l_{b}+(\rho+p)(l_{a}n_{b}+n_{a}l_{b})+p g_{ab}
\end{equation}

In the comoving co-ordinates
($v,r,\theta_{1},\theta_{2},...,\theta_{n}$), the two eigen
vectors of energy-momentum tensor namely $l_{a}$ and $n_{a}$ are
linearly independent future pointing light-like vectors (null
vectors) having components $l_{a}=(1,0,0,...,0)$ and
$n_{a}=(f/2,-1,0,...,0)$ and they satisfy the relations

\begin{equation}
l_{a}l^{a}=n_{a}n^{a}=0,~ l_{a}n^{a}=-1
\end{equation}

Here, $\mu$ is the energy density of Vaidya null radiation, $\rho$
is the energy density and $p$ is the radial pressure satisfying
$p=-\sigma\rho$ with $\rho= C(v)r^{-n(1-\sigma)}$. The density
$\rho$ and pressure $p$ come from the Lovelock gravity. The
solution (13) is called the generalized Vaidya solution in
Lovelock theory. Clearly this equation denotes the linear barotropic
equation of state. We must follow that when $\sigma=-\frac{1}{3}$ the
radiation era is signified whereas for $\sigma=0$  pressureless
dust filled model is realised. When $\sigma>\frac{1}{3}$, it denotes dark energy, to be very particular
quintessence era. At last when $\sigma=1$ it implies  $\Lambda$CDM whereas $\sigma>1$ means phantom era.\\

\section{Gravitational Collapse in GB Gravity}

Setting $p=2,c_{0}=0,c_{1}=1$ and $c_{2}=\alpha$, the generalized
Vaidya solution (13) in Lovelock gravity reduces to generalized
Vaidya solution in GB gravity. Here the parameter $\alpha$ is
called the GB coupling parameter having dimension $(length)^{2}$
and is related to string tension as $\alpha^{-1}$.In the GB theory
the function $f(v,~r)$ can be expressed as [Cai et al, 2008]

\begin{equation}\label{2}
f(v,~r)=1+\frac{1}{2\left(n-1\right)\left(n-2\right)\alpha}\left[r^{2}
\pm \sqrt{r^{4} +\frac{64\pi G
\left(n-1\right)\left(n-2\right)\alpha}{n}\left(\frac{m(v)}{\Omega_{n}r^{n-3}}
+\frac{C(v)\Theta(r)}{r^{n-3}}\right)}\right]
\end{equation}

We shall discuss the existence of NS in generalized
Vaidya space-time by studying radial null geodesics. In fact, we
shall examine whether it is possible to have outgoing radial null
geodesics which were terminated in the past at the central
singularity $r=0$. The nature of the singularity (NS or BH) can be characterized by the existence
of radial null geodesics emerging from the singularity. The
singularity is at least locally naked if there exist such
geodesics and if no such geodesics exist it is a BH.\\

The equation for outgoing radial null geodesics can be obtained
from equation (\ref{3}) by putting $ds^{2}=0$ and
$d\Omega_{n}^{2}=0$ as

\begin{equation}\label{4}
\frac{dv}{dr}=\frac{2}{f(v,~r)}.
\end{equation}

It can be seen easily that $r=0,~v=0$ corresponds to a singularity
of the above differential equation. Suppose $X=\frac{v}{r}$ then
we shall study the limiting behaviour of the function $X$ as we
approach the singularity at $r=0,~v=0$ along the radial null
geodesic. If we denote the limiting value by $X_{0}$ then

\begin{eqnarray}
\begin{array}{c}
X_{0}\\\\
{}
\end{array}
\begin{array}{c}
=lim~~ X \\
\begin{tiny}v\rightarrow 0\end{tiny}\\
\begin{tiny}r\rightarrow 0\end{tiny}
\end{array}
\begin{array}{c}
=lim~~ \frac{v}{r} \\
\begin{tiny}v\rightarrow 0\end{tiny}\\
\begin{tiny}r\rightarrow 0\end{tiny}
\end{array}
\begin{array}{c}
=lim~~ \frac{dv}{dr} \\
\begin{tiny}v\rightarrow 0\end{tiny}\\
\begin{tiny}r\rightarrow 0\end{tiny}
\end{array}
\begin{array}{c}
=~~~~lim~~ \frac{2}{f(v,~r)} \\
\begin{tiny}v\rightarrow 0\end{tiny}\\
\begin{tiny}r\rightarrow 0\end{tiny}
 {}
\end{array}
\end{eqnarray}

Using (20) and (22), we have

\begin{eqnarray}\label{6a}
 X_{0}=
\frac{2}{\begin{array}llim\\
\begin{tiny}v\rightarrow 0\end{tiny}\\
\begin{tiny}r\rightarrow 0\end{tiny}
\end{array}
\left\{1+\frac{1}{2\left(n-1\right)\left(n-2\right)\alpha}\left[r^{2}
\pm \sqrt{r^{4}+\frac{64\pi G
\left(n-1\right)\left(n-2\right)\alpha}{n}\left(\frac{m(v)}{\Omega_{n}r^{n-3}}
+\frac{C(V)\Theta(r)}{r^{n-3}}\right)}\right]\right\}}
\end{eqnarray}

Now choosing $m(v)=m_{0}v^{n-3}$ and
${C}(v)={C}_{0}v^{n(1-\sigma)-4}$, the equation (23) yields

\begin{equation}\label{7}
X_{0}=\frac{2}{1\pm\frac{1}{2(n-1)(n-2)\alpha}\left[\sqrt{
\left(\frac{64\pi G(n-1)(n-2)\alpha}{n} \right)
\left(\frac{m_{0}X_{0}^{n-3}\Gamma\left(1+\frac{n}{2} \right)
}{\pi^{\frac{n}{2}}} +\frac{C_{0}X_{0}^{n(1-\sigma)-4}}{n\sigma+1}
\right)} \right]}
\end{equation}

On simplifying we have the algebraic equation of $X_{0}$ as

\begin{equation}\label{8}
\frac{16\pi^{1-\frac{n}{2}}G
m_{0}\Gamma\left(1+\frac{n}{2}\right)}{n\alpha\left(n-1\right)
\left(n-2\right)}X_{0}^{n-1}+\frac{16\pi G
C_{0}}{n\alpha\left(n-1\right)\left(n-2\right)
\left(n\sigma+1\right)}X_{0}^{n\left(1-\sigma\right)-2}-X_{0}^{2}+4X_{0}-4=0
\end{equation}\\

This is a very complicated algebraic equation. So as it will be
bit difficult to express the value of the root $X_{0}$ in terms of
all the parameters, we will vary the value of these parameters to
check whether the concerned equation gives positive real root for
these set of parametric values. Below we have prepared a table I
where some particular sets of values of the parameters have been
considered. At first we have to note that as we increase the
dimension the tendency of having positive root increases. Besides
as the value of $m_{0}$ is been increased for lower dimension we
may not have positive root. So in this case we will have a BH
singularity. The parameter $\alpha$ takes a leading role in having
positive roots. If we take smaller values of $\alpha$ the tendency
of `not having' positive roots increases even if for larger
dimensions too. $\alpha$ plays a role of multiplier to the GB term
$R_{GB}(=R^{\alpha\beta\gamma\delta}R_{\alpha\beta\gamma\delta}-4R^{\alpha\beta}R_{\alpha\beta}+R^{2})$)
in the action [Boulware et al, 1985]. When $\alpha$ is very low it
is very obvious that the impact of the GB term in the gravity will
be lower. As $\alpha\rightarrow 0$ we get back the Einstein
gravity. So we can predict that tendency of formation of BH
reduces due to expanding nature of universe. Hence physically we
can interpret this phenomenon as : in Einstein gravity
circumstances were ready to give a BH or to wrap a singularity by
as event horizon. On the contrary when we introduce the effect of
GB term, i.e., we take the expanding universe in account it is
observed that the event horizon is not been formed under the same
circumstances. The negative pressure here is forcing the
singularity to be naked. $C_{0}$ has no such impact on getting
positive roots.\\

At last we will discuss the impact of the most important parameter
$\sigma$ which is actually determining the nature of the fluid
present inside the universe. When $\sigma>\frac{1}{3}$ we know it
denotes the dark energy (DE) era (both quintessence and phantom)
as this will not obey the strong energy condition ($\rho+3p>0$).
For this case we have seen that the chance of getting a BH is less
in the higher dimensions. In some previous works when Debnath et
al (2003, 2004), Banerjee et al (2003) have studied the collapse
in higher dimensional Tolman-Bondi model they have shown that in
Einstein gravity the possibility of getting a BH increases in
higher dimensions. So we have reached an opposite result in our
current work i.e., when GB gravity is considered. This may  be a
strong impact of negative pressure of the fluid.\\

So ultimately we may tell that the possibility of the formation of
BH in expanding universe is bleak. In literature, Babichev et al
(2004) and Biswas et al (2011) had shown that the accretion of
dark fluid either reduces the mass of the BH or weakens its
accretion procedure. Similarly, here we can conclude that while
forming a singularity by collapsing, DE/expanding universe opposes
the formation of the BH and rather it increases
the preferences of growing a NS.\\

\begin{center}
\begin{tabular}{|l|}
\hline\hline
~~$m_{0}$~~~~~~~~~~~~$\sigma$~~~~~~~~~~~~~~~~~~~~~$\alpha$~~~~~~~~~~~${C}_{0}$
~~~~~~~~~~~~~~~~~~~~~~~~~~Positive roots ($X_{0}$) \\ \hline
\\
~~~~~~~~~~~~~~~~~~~~~~~~~~~~~~~~~~~~~~~~~~~~~~~~~~~~~~~~~~~~~~~~~~~~~~~~5D~~~~~~~~~~~~6D~~~~~~~~~~7D~~~~~~~~~8D~~~~~~~~~9D~~~~~~~~~10D
\\ \hline\hline
\\
~0.01~~~~~~~~~2(phantom)~~~~~~~~~~0.2~~~~~~~~0.5~~~~~~~~~~~~2.93572~~~~~~2.61882~~~0.73852~~~0.68112~~~~0.65449~~~~0.64221
\\
~0.01~~~~~~~~~2(phantom)~~~~~~~~0.002~~~~~~0.0002~~~~~~~~~~~$-$~~~~~~~~~~~0.84339~~~1.03722~~~1.13178~~~~1.17769~~~~1.19622
\\
~0.01~~~~~~~~~2(phantom)~~~~~~0.00002~~~~~~0.2~~~~~~~~~~~~~~$-$~~~~~~~~~~~~~~$-$~~~~~~~~~~~$-$~~~~~~~~~~~$-$~~~~~~~~~$-$~~~~~~~~~~~~$-$
\\\hline\hline\\
~0.25~~~~~~~~~1($\Lambda$CDM)~~~~~~~~~0.0003~~~~~~~0.05~~~~~~~~~~~~~$-$~~~~~~~~~~~~~$-$~~~~~~~~~~~$-$~~~~~~~~~~~$-$~~~~~~~~~$-$~~~~~~~~~~~~$-$
\\
~~0.5~~~~~~~~~1($\Lambda$CDM)~~~~~~~~~~~~0.3~~~~~~~~~0.5~~~~~~~~~~~~~~~$-$~~~~~~~~~~0.78632~~~~1.17709~~~1.27789~~~1.31614~~~1.32483
\\
~~0.5~~~~~~~~~1($\Lambda$CDM)~~~~~~~~~~~~0.3~~~~~~~~~0.5~~~~~~~~~~~~~~~$-$~~~~~~~~~~~~~$-$~~~~~~~~~~~$-$~~~~~~~~~~$-$~~~~~~~~0.93492~~~1.01816
\\
~0.75~~~~~~~~~1($\Lambda$CDM)~~~~~~~~~~~0.3~~~~~~~~~0.05~~~~~~~~~~~~~~$-$~~~~~~~~~~0.73861~~~~1.05256~~~1.18076~~~1.23866~~~1.26091
\\
~0.75~~~~~~~~~1($\Lambda$CDM)~~~~~~~~~0.0003~~~~~~0.05~~~~~~~~~~~~~~~$-$~~~~~~~~~~~~$-$~~~~~~~~~~~$-$~~~~~~~~~~~$-$~~~~~~~~~$-$~~~~~~~~~~~~$-$
\\\hline\hline\\
0.25~~~~~~~~~0.5(quintessence)~~~~0.3~~~~~~~~0.5~~~~~~~~~~~~~~~$-$~~~~~~~~~~0.83101~~~~1.17329~~~1.30484~~~1.36071~~~1.37884
\\
~0.25~~~~~~~~0.5(quintessence)~~0.003~~~~~~~0.5~~~~~~~~~~~~~~~$-$~~~~~~~~~~~~$-$~~~~~~~~~0.009856~~0.18802~~~0.42496~~~0.60799
\\
~~0.5~~~~~~~~0.5(quintessence)~~~~0.01~~~~~~~0.5~~~~~~~~~~~~~~~$-$~~~~~~~~~~~~$-$~~~~~~~~~0.09242~~~0.44689~~~0.68008~~~0.81153
\\
~0.75~~~~~~~~0.5(quintessence)~~~~0.3~~~~~~~~0.5~~~~~~~~~~~~~~~$-$~~~~~~~~~~0.73861~~~1.05256~~~1.18076~~~1.23866~~~1.26091
\\\hline\hline\\
~0.2~~~~~~~~~0.3~~~~~~~~~~~~~~~~~~~~~~~0.2~~~~~~~~0.5~~~~~~~~~~~~0.00003~~~~~0.6537~~~~~1.00079~~~1.16498~~~1.266~~~~~~2.1094
\\
~0.2~~~~~~~~~0.1~~~~~~~~~~~~~~~~~~~~~~~0.2~~~~~~~~0.5~~~~~~~~~~~~0.11429~~~~~0.6280~~~~~0.9057~~~~~1.05539~~~1.099~~~~~~1.1453
\\\hline\hline\\
~0.2~~~~~~~~0(dust)~~~~~~~~~~~~~~~~~~0.2~~~~~~~~0.5~~~~~~~~~~~~0.159342~~~0.598256~~0.843498~~0.980193~~1.0618~~~~1.11264
\\\hline\hline\\
~0.2~~~~~~$-\frac{1}{3}$(radiation)~~~~~~~~0.0002~~~~~~~0.5~~~~~~~~~~~~~~$-$~~~~~~~~~~~~~~$-$~~~~~~~~~~~$-$~~~~~~~~~~~$-$~~~~~~~~~$-$~~~~~~~~~~~~$-$
\\
\\ \hline\hline
\end{tabular}
\end{center}

{\bf Table I:} Nature of the roots ($X_{0}$) of the equation (25) for various values of parameters involved.\\

\section{Gravitational Collapse in Dimensionally Continued Lovelock \\ Gravity}

If we consider dimensionally continued Lovelock gravity, the
solution of the equation (13) yields [Cai et al, 2008; Ilha et al,
1997, 1999; Nozawa, 2006]

\begin{equation}\label{9}
f(v,~r)=\left\{
\begin{array}{cccc}
1-\left[\frac{m(v)+\Omega_{n}C(v)\Theta(r)}{r}\right]^{\frac{2}{n}}+\frac{16
\pi G r^{2}}{\Omega_{n}n!}~~,~~~~for~ even~ values~  of~ n.\\\\
1-\left[m(v)+\Omega_{n}C(v)\Theta(r)\right]^{\frac{2}{n+1}}+\frac{256\pi^{2}G^{2}r^{2}}
{\Omega_{n}^{2}(n!)^{2}}~~,~~~~for~odd~ values~  of~ n.
\end{array}
\right.
\end{equation}

In the first solution of equation (\ref{9}), put $m(v)=m_{0}v$ and
$C(v)=C_{0}v^{-n\sigma}$ and using (22), we have

\begin{equation}\label{11}
 X_{0}
=\frac{2}{1-\left[X_{0}m_{0}+\frac{\Omega_{n}}{n\sigma+1}\frac{1}{X_{0}^{n\sigma}}C_{0}\right]^{\frac{2}{n}}}
\end{equation}

On simplifying we have the algebraic equation of $X_{0}$ for even
values of $n$ as

\begin{equation}\label{12}
X_{0}-2-X_{0}\left[m_{0}X_{0}+\frac{\pi^{\frac{n}{2}}}{\Gamma\left(1+\frac{n}{2}\right)
\left(n\sigma+1\right)}\frac{C_{0}}{X_{0}^{n\sigma}}\right]^{\frac{2}{n}}=0
\end{equation}

In the second solution of equation (\ref{9}), put $m(v)=m_{0}v$ and
$C(v)=C_{0}v^{-n\sigma-1}$ and using (22) and applying the limits,
we get the algebraic equation of $X_{0}$ for odd values of $n$ as

\begin{equation}\label{13}
X_{0}-2-X_{0}\left[\frac{\pi^{\frac{n}{2}}}{\Gamma\left(1+\frac{n}{2}\right)
\left(n\sigma+1\right)}\frac{C_{0}}{X_{0}^{n\sigma+1}}\right]^{\frac{2}{n+1}}=0
\end{equation}

Like the previous section here also it is tough to determine the
explicit form of the solution in terms of the different governing
parameters. So here also we have varied the values of the
parameters and dimension to observe their impact on forming
positive or non-positive/non-real solution. Like the previous
section here also we have listed all the data in a table formate
(table II). Now analyzing the table we can say that $\sigma$ has
no big impact on the nature of root. But if $m_{0}$ and $C_{0}$
are both small then irrespective of the dimension we will get
positive roots. But as either of these two parameters are
increased we will have  real positive solution only if the
dimension is odd. So in this case, the singularity will be naked.
For even dimension we will have
black holes.\\

It is to be noted that when $\sigma=2$, i.e., in phantom era no BH
forms. As like the previous section we interpret this phenomenon
as follows : the negative pressure of DE is opposing a
singularity to be a BH and forcing it to be a NS.\\

\begin{center}
\begin{tabular}{|l|}
\hline\hline
~~$m_{0}$~~~~~~~~~~~$\sigma$~~~~~~~~~~~~~~~~~~~~~${C}_{0}$
~~~~~~~~~~~~~~~~~~~~~~~~~~~~~~~~~Positive roots ($X_{0}$) \\
\hline
\\
~~~~~~~~~~~~~~~~~~~~~~~~~~~~~~~~~~~~~~~~~~~~~~~~~~~~~~~~~~4D~~~~~~~~~~~5D~~~~~~~~~~~~6D~~~~~~~~~~~~7D~~~~~~~~~~~8D~~~~~~~~~~~9D
\\ \hline\hline
\\
~~0.01~~~~~~~2(phantom)~~~~~~~~0.001~~~~~~~~~~2.04176~~~~~~2.0043~~~~~~~2.36332~~~~~~2.01212~~~~~~2.88472~~~~~2.01929
\\
\hline\hline
\\
~~0.25~~~~~~~1($\Lambda$CDM)~~~~~~0.000002~~~~~~~~~~~~~$-$~~~~~~~~~~2.00072~~~~~~~~~$-$~~~~~~~~~~2.00601~~~~~~~~~~$-$~~~~~~~~2.01635
\\
~~0.75~~~~~~~1($\Lambda$CDM)~~~~~~0.000002~~~~~~~~~~~~~$-$~~~~~~~~~~2.00072~~~~~~~~~$-$~~~~~~~~~~2.00601~~~~~~~~~~$-$~~~~~~~~2.01635
\\\hline\hline
\\
~~0.25~~~~~~0.5(quintessence)~~0.0002~~~~~~~~~~~$-$~~~~~~~~~~2.01536~~~~~~~~~~$-$~~~~~~~~~~2.0594~~~~~~~~~~~$-$~~~~~~~~2.10965
\\
~~0.01~~~~~~0.5(quintessence)~~0.1~~~~~~~~~~~~2.20573~~~~~~2.33127~~~~~~2.57982~~~~~~2.45778~~~~~2.96694~~~~~2.50744
\\
~~0.75~~~~~~0.5(quintessence)~~0.0002~~~~~~~~~~~$-$~~~~~~~~~~2.01536~~~~~~~~~~$-$~~~~~~~~~2.0594~~~~~~~~~~~~$-$~~~~~~~2.10965
\\\hline\hline\\
~~0.01~~~~~~0.25~~~~~~~~~~~~~~~~~~~~~2~~~~~~~~~~~~~~~~$-$~~~~~~~~~~4.65149~~~~~~~9.3907~~~~~~4.42596~~~~~~~6.3532~~~~~4.11906
\\
~~0.25~~~~~~0.25~~~~~~~~~~~~~~~~~~~~~2~~~~~~~~~~~~~~~~$-$~~~~~~~~~~4.65149~~~~~~~~~~$-$~~~~~~~~~4.42596~~~~~~~~~~~$-$~~~~~~~4.11906
\\
~~0.75~~~~~~0.25~~~~~~~~~~~~~~~~~~~~~2~~~~~~~~~~~~~~~~$-$~~~~~~~~~~4.65149~~~~~~~~~~$-$~~~~~~~~~4.42596~~~~~~~~~~~$-$~~~~~~~4.11906
\\\hline\hline\\
~~0.75~~~~~~0(dust)~~~~~~~~~~~~~~~~~2~~~~~~~~~~~~~~~$-$~~~~~~~~~~~2.04135~~~~~~~~~~$-$~~~~~~~~~2.17054~~~~~~~~~~~$-$~~~~~~~2.33073
\\\hline\hline\\
~~0.75~~~~$-\frac{1}{3}$(radiation)~~~~~~~~~~2~~~~~~~~~~~~~~~$-$~~~~~~~~~~~~~~$-$~~~~~~~~~~~~~~$-$~~~~~~~~~~~~$-$~~~~~~~~~~~~~~~~$-$~~~~~~~~~$-$
\\
\\ \hline\hline
\end{tabular}
\end{center}

{\bf Table II:} Nature of the roots ($X_{0}$) of the equations (28) and (29)
for various values of parameters involved.\\

\section{Gravitational Collapse in Pure Lovelock Gravity}

In pure Lovelock gravity, only two coefficients $c_{0}$ and
$c_{k}$ are non-vanishing with $1\leq k\leq
\left[\frac{n+1}{2}\right]$. Choosing
$c_{k}=\frac{(n-2k+1)!}{(n+1)!}~\alpha^{2k-2}$, the solution of
$f(v,~r)$ can be found from (13) and is given by [Cai et al, 2008]

\begin{equation}\label{14}
\alpha^{2k-2}\left\{1-f(v,~r)\right\}^{k}=-\frac{c_{0}r^{2k}}{n(n+1)}+\frac{16\pi
G}{n}\left(\frac{m(v)}{\Omega_{n}r^{n-2k+1}}+\frac{C(v)\Theta(r)}{r^{n-2k+1}}\right)
\end{equation}

where $\alpha$ is a constant length scale.\\

Assuming $m(v)=m_{0}v^{n-2k+1}$ and $C(v)=C_{0}v^{n-2k-n\sigma}$,
using (22) and applying the limits we have,

\begin{equation}\label{17}
\left\{1-\frac{2}{X_{0}}\right\}^{k}\alpha^{2k-2}=\frac{16\pi
G}{n}\left\{\frac{m_{0}\Gamma
\left(1+\frac{n}{2}\right)X_{0}^{n-2k+1}}{\pi^{\frac{n}{2}}}+\frac{C_{0}X_{0}^{n-2k-n\sigma}}{n\sigma+1}\right\}\\
\end{equation}

and simplifying, we get the algebraic equation of $X_{0}$ as

\begin{equation}\label{18}
16\pi^{1-\frac{n}{2}}G~\Gamma\left(1+\frac{n}{2}\right)m_{0}X_{0}^{n-2k+1}+\frac{16\pi
G C_{0}}
{n\sigma+1}X_{0}^{n-2k-n\sigma}-\left(1-\frac{2}{X_{0}}\right)^{k}n\alpha^{2k-2}=0
\end{equation}

As this is also a complicated equation to give an explicit
solution of $X_{0}$ as a function of the parameters and dimension,
we will again check the values of roots extracted from the
equation for constant values of the parameters. All the parametric
values, dimensions and the roots found are given below in a
tabular form (table III). Here we can see unlike the previous
cases irrespective of whatever parametric values used we are
having non-positive roots in each cases. It implies that in pure
Lovelock gravity we will always have a black hole.\\

\begin{center}
\begin{tabular}{|l|}
\hline\hline
~~$m_{0}$~~~~~~~~~~$\sigma$~~~~~~~~~~~~~~~~~~~~~${C}_{0}$
~~~~~~~~~~~~~~~~~~~~~~~~~~~~~~~~~~~~~Positive roots ($X_{0}$) \\
\hline
\\
~~~~~~~~~~~~~~~~~~~~~~~~~~~~~~~~~~~~~~~~~~~~~~~~~~~~~~~~~~~~4D~~~~~~~~~~~~5D~~~~~~~~~~~~~~~6D~~~~~~~~~~~7D~~~~~~~~~~~8D~~~~~~~~~9D
\\ \hline\hline
\\

~~0.11~~~~~2(phantom)~~~~~~~~~0.001~~~~~~~~~~~~~~~~~$-$~~~~~~~~~~~~~$-$~~~~~~~~~~~~~~~~$-$~~~~~~~~~~~~$-$~~~~~~~~~~~~~$-$~~~~~~~~~~~$-$
\\
~~1~~~~~~~~~2(phantom)~~~~~~~~~0.001~~~~~~~~~~~~~~~~~$-$~~~~~~~~~~~~~$-$~~~~~~~~~~~~~~~~$-$~~~~~~~~~~~~$-$~~~~~~~~~~~~~$-$~~~~~~~~~~~$-$
\\
~~0.3~~~~~~~2(phantom)~~~~~~~~~0.001~~~~~~~~~~~~~~~~~$-$~~~~~~~~~~~~~$-$~~~~~~~~~~~~~~~~$-$~~~~~~~~~~~~$-$~~~~~~~~~~~~~$-$~~~~~~~~~~~$-$
\\
~~0.3~~~~~.75(quintessence)~~~~~~~~~1~~~~~~~~~~~~~~~~~$-$~~~~~~~~~~~~~$-$~~~~~~~~~~~~~~~~$-$~~~~~~~~~~~~$-$~~~~~~~~~~~~~$-$~~~~~~~~~~~$-$
\\
~~0.11~~~~.75(quintessence)~~~~~~~~1~~~~~~~~~~~~~~~~~$-$~~~~~~~~~~~~~$-$~~~~~~~~~~~~~~~~$-$~~~~~~~~~~~~$-$~~~~~~~~~~~~~$-$~~~~~~~~~~~$-$
\\
~~1~~~~~~~.75(quintessence)~~~~~~~~~1~~~~~~~~~~~~~~~~~$-$~~~~~~~~~~~~~$-$~~~~~~~~~~~~~~~~$-$~~~~~~~~~~~~$-$~~~~~~~~~~~~~$-$~~~~~~~~~~~$-$
\\

~~0.3~~~~~0.2~~~~~~~~~~~~~~~~~~~~~~~~~~~2~~~~~~~~~~~~~~~~~~$-$~~~~~~~~~~~~~$-$~~~~~~~~~~~~~~~~$-$~~~~~~~~~~~~$-$~~~~~~~~~~~~~$-$~~~~~~~~~~~$-$
\\
~~0.11~~~~0.2~~~~~~~~~~~~~~~~~~~~~~~~~~2~~~~~~~~~~~~~~~~~~$-$~~~~~~~~~~~~~$-$~~~~~~~~~~~~~~~~$-$~~~~~~~~~~~~$-$~~~~~~~~~~~~~$-$~~~~~~~~~~~$-$
\\
~~1~~~~~~~0.2~~~~~~~~~~~~~~~~~~~~~~~~~~~2~~~~~~~~~~~~~~~~~~$-$~~~~~~~~~~~~~$-$~~~~~~~~~~~~~~~~$-$~~~~~~~~~~~~$-$~~~~~~~~~~~~~$-$~~~~~~~~~~~$-$
\\
~~1~~~~~~~0(dust)~~~~~~~~~~~~~~~~~~~~~2~~~~~~~~~~~~~~~~~~$-$~~~~~~~~~~~~~$-$~~~~~~~~~~~~~~~~$-$~~~~~~~~~~~~$-$~~~~~~~~~~~~~$-$~~~~~~~~~~~$-$
\\
~~1~~~~~~-$\frac{1}{3}$(radiation)
~~~~~~~~~~~~~~2~~~~~~~~~~~~~~~~~~$-$~~~~~~~~~~~~~$-$~~~~~~~~~~~~~~~~$-$~~~~~~~~~~~~$-$~~~~~~~~~~~~~$-$~~~~~~~~~~~$-$

\\ \hline\hline
\end{tabular}
\end{center}

{\bf Table III:} Nature of the roots ($X_{0}$) of the equation (32) for various values of parameters involved.\\

\section{Gravitational collapse when some fluid is also accreting upon the collapsing object}

It is an obvious fact that a highly massive star accretes fluid
around it mainly from the dust cloud in which the star is present
or from any companion star which is present in a binary system
with the super massive star as a companion. As the star absorbs
mass the gravitational pull increases irrespective of the
increment of the electronic force at its surface. So when the
system collapses under its own gravity there may be an impact of
the accreting fluid. Even when the accreting fluid is of DE type
this may cause a chance of evolving a NS whereas the fluid like
normal matter may increase the chance of forming an event horizon.
In this section we will check whether our speculation is true or
not.\\

It is well known that the rate of change of mass of the central
object for accreting phenomena is $\Omega_{n}r^{n}T_{0}^{1}$
(generalizing the result of Babichev et al (2004)). So the total
mass will be
$$M-\dot{M}dv=M-\Omega_{n}r^{n}T_{0}^{1}dv$$
Here, $T_{0}^{1}$ is non-diagonal stress energy tensor component
evolved due to the accreting fluid given by (16). So equation (20)
will be changed to
\begin{equation}\label{}
\bar{f}(v, ~r)=f(v,~r)+\frac{\Omega_{n}}{r^{n}}\left[\frac{\dot{m}(v)}{\Omega_{n}}+\dot{C}(v)\Theta(r)\right]
\end{equation}
and equation (22) becomes
\begin{eqnarray}\label{}
 X_{0}=
\frac{2}{\begin{array}llim\\
\begin{tiny}v\rightarrow 0\end{tiny}\\
\begin{tiny}r\rightarrow 0\end{tiny}
\end{array}
\left[f(v,~r)+\frac{\Omega_{n}}{r^{n}}\left\{\frac{\dot{m}(v)}{\Omega_{n}}+\dot{C}(v)\Theta(r)\right\}
\right]}
\end{eqnarray}

where $f(v,r)$ is given in eq.(20).\\

For all gravity, we take $m(v)=\bar{m}_{0}v^{n+1}$ and
$C(v)=\bar{C}_{0}v^{n(1-\sigma)}$ we get

\begin{equation}
\bar{m}_{0}(n+1)X_{0}^{n+1}+\frac{\bar{C}_{0}\pi^{\frac{n}{2}}n(1-\sigma)}{(n\sigma+1)\Gamma(1+\frac{n}{2})}~X_{0}^{n(1-\sigma)}+X_{0}-2=0
\end{equation}

Clearly in the equation (35), $X_{0}$ is not depending at all upon
the GB parameter $\alpha$ or any other information of the Vaidya
metric given by eq.(20). So if we try to modify eqs.(28), (29) and
(32) we will have the same equation (35). While the fluid is
accreting upon a collapsing object, the nature of singularity (NS
or BH) will be same for above mentioned three types of
gravities.\\

Like the previous sections it is really tough to get the explicit
solution of $X_{0}$ as a function of parameters and dimension. So
we will again check the values of roots extracted from the
equation for constant values of parameters. All the parametric
values, dimensions and the roots found are given in a
tabular form (table IV).\\

Here we have seen for all the ranges of different parameters,
irrespective of dimensions we get NS. As we have stated before in
2004, Babichev et al have shown that while our universe is
expanding the existence of highly compact objects like BH at Big
Rip is quite impossible. They have shown that the fluid (DE) which
is responsible for creating such a negative pressure and
responsible for the accelerated expansion of present day universe
decreases the mass of BH while it is been accreted upon the BH
(such that before arriving the Big Rip point all the BHs will be
evaporated completely). Likewise when we have assumed the fluid is
accreted upon a collapsing object it will be very obvious for the
fluid that it will oppose the collapsing object to have an event
horizon (i.e., to form a BH) and will have a tendency to produce a
NS.\\

\begin{center}
\begin{tabular}{|l|}
\hline\hline
~~$\bar{m}_{0}$~~~~~~~~~~~~$\sigma$~~~~~~~~~~~~~~~~~~${\bar{C}}_{0}$
~~~~~~~~~~~~~~~~~~~~~~~~~~~~~~~~~~~~~~~~~Positive roots ($X_{0}$) \\
\hline
\\
~~~~~~~~~~~~~~~~~~~~~~~~~~~~~~~~~~~~~~~~~~~~~~~~~~~~~~~~~~~~4D~~~~~~~~~~5D~~~~~~~~~~~~~6D~~~~~~~~~~~~~7D~~~~~~~~~~~~~~8D~~~~~~~~~~~~~~9D
\\ \hline\hline
\\
~~0.2~~~~~~~~~2(phantom)~~~~~~~~0.5~~~~~~~~~~~~~~~1.2455~~~~~~1.1555~~~~~~~~~1.1034~~~~~~~~1.0688~~~~~~~~~~1.0431~~~~~~~~~~1.022
\\
~~2~~~~~~~~~~~2(phantom)~~~~~~0.005~~~~~~~~~~~~~~0.6157~~~~~~~0.645~~~~~~~~~~0.6729~~~~~~~~0.6969~~~~~~~~~~~0.717~~~~~~~~~~0.7348
\\
~~0.002~~~~~~1($\Lambda$CDM)~~~~~~~~0.005~~~~~~~~~~~~~~1.955~~~~~~~~1.896~~~~~~~~~~1.807~~~~~~~~~~1.7051~~~~~~~~~1.6090~~~~~~~~~~1.5270
\\
~~0.2~~~~~~~~~1($\Lambda$CDM)~~~~~~~~~~~5~~~~~~~~~~~~~~~~1.1312~~~~~~1.0452~~~~~~~~~~1.00~~~~~~~~~~~0.9741~~~~~~~~~0.9586~~~~~~~~~~0.9488
\\\hline\hline
\\
~~0.2~~~~~~~~~0.5(quintessence)~~5~~~~~~~~~~~~~~~~0.2251~~~~~~0.2665~~~~~~~~~~0.3193~~~~~~~~0.3752~~~~~~~~~~0.4321~~~~~~~~~0.4892
\\
~~0.02~~~~~~~0.5(quintessence)~~0.05~~~~~~~~~~~~1.9408~~~~~~~1.8700~~~~~~~~~~1.7730~~~~~~~~1.579~~~~~~~~~~~~1.52~~~~~~~~~~~1.5041
\\
~~0.01~~~~~~~0.25~~~~~~~~~~~~~~~~~~0.005~~~~~~~~~~~1.7902~~~~~~~~1.6337~~~~~~~~~1.4981~~~~~~~~~1.398~~~~~~~~~~~1.326~~~~~~~~~~1.2759
\\
~~2.5~~~~~~~~~0.1~~~~~~~~~~~~~~~~~~~~0.5~~~~~~~~~~~~~0.4871~~~~~~~~~0.512~~~~~~~~~~0.5570~~~~~~~~0.5932~~~~~~~~~~0.6273~~~~~~~~0.6677
\\
\hline\hline
\end{tabular}
\end{center}

{\bf Table IV:} Nature of the roots ($X_{0}$) of the equation (35) for various values of parameters involved.\\

\section{Conclusions}

The generalized Vaidya solution in Lovelock theory of gravity in
$(n+2)$-dimensions have been thoroughly assumed in this work.
Gauss-Bonnet gravity, dimensionally continued Lovelock gravity and
pure Lovelock gravity have been considered and it has been
successfully shown that these three particular forms of Lovelock
theory of gravity can be constructed by suitable choice of
parameters. We have studied the occurrence of singularities formed
by the gravitational collapse in the above three particular forms
of Lovelock theory of gravity.\\

In GB gravity three major remarkable results have been observed :
(i) Controlling the parameter $\alpha$ when we go back to Einstein
gravity we find that the possibility of forming a NS increases.
(ii) When we increase dimensions possibility of finding NS
increases while for Einstein gravity contrary to the increase in
chances of forming NS, the BH is formed for higher dimensions.
(iii) The equation of state parameter ($\sigma$ here), when
denotes negative pressure, tendency of forming NS increases. From
the above three observations we can infer that for modified
gravity theory chances of forming naked singularity is much much
higher than forming a BH. Dimensionally continued Lovelock gravity
theory resembles  with the GB gravity in the matter of collapse
whereas pure Lovelock gravity allows only BH to be formed. We can
interpret this as, the gravity in case of pure Lovelock gravity
theory is more stronger than the expanding force (dark force). At
last in section 6 we have considered collapse under accretion
where we have seen it is almost impossible to form a BH while the
accretion is going upon the collapsing object in the expanding
universe (presented by the modified gravity theory). Babichev et
al (2004) have shown that under the accretion of phantom fluid the
BH mass gets evaporated and the concerned BH will never be able to
face the Big Rip. But here in the current work we can speculate
that in expanding universe (specially in phantom era) no BH at all
will be formed if we are in Gauss-Bonnet gravity or considering
accretion procedure upon the collapsing object.\\

\begin{contribution}

{\bf Acknowledgement :}\\

RB is thankful to West Bengal State Govt for awarding JRF.

\end{contribution}

\end{document}